\documentclass[printer]{aa}
\usepackage{graphics,epsfig,lscape}
\usepackage{natbib,aas_macros}
\bibpunct{(}{)}{;}{a}{}{,}

\def\cm-2{cm$^{-2}$}

\def\n6217{\object{NGC~6217}}
\newcommand{\ergcm}[1]{$\cdot 10^{#1}$ \hbox{erg cm$^{-2}$ s$^{-1}$}}
\newcommand{\oergcm}[1]{$10^{#1}$ erg cm$^{-2}$ s$^{-1}$}
\newcommand{\ergs}[1]{$\cdot 10^{#1}$ \hbox{erg s$^{-1}$}}

\newcommand{\hcm}[1]{$\cdot 10^{#1}$ cm$^{-2}$}

\newcommand{\nh}{\hbox{$N_{\rm H}$}}

\newcommand{\ct}{ct s$^{-1}$}

\newcommand{\perr}{r$_{90}$}
\newcommand{\exil}{$ML_{\rm exi}$}
\begin{document}

   \title{A possible X-ray jet from the starburst galaxy \n6217}

   \author{W. Pietsch\inst{1} \and  
           H. Arp\inst{2} 
          }
\institute{Max-Planck-Institut f\"ur extraterrestrische Physik, 85741 Garching, Germany \and
           Max-Planck-Institut f\"ur Astrophysik, 85741 Garching, Germany 
          }
     
     \offprints{W. Pietsch (wnp@mpe.mpg.de)}

         \date{Received; accepted }

        \abstract{
     Deep ROSAT PSPC and HRI observations of the nuclear starburst, UV 
flat spectrum spiral \n6217 reveal a jet-like
X-ray filament extending out 2\farcm7 (18.8 kpc) to the SW of the galaxy. 
Radio images of \n6217 show an extent in the same direction, giving further
evidence for a one-sided X-ray jet interpretation of this feature. 
It's X-ray spectrum is harder than that of \n6217 and the 
luminosity in the 0.5--2.0 keV band $\sim$1.7\ergs{39}. We compare our findings 
to parameters of other X-ray jets from other active galaxies.
We also give positions of a total of 91 X-ray sources detected in the field 
to a limiting 0.5--2.0 keV flux of $\sim 2.2$\ergcm{-15} and propose
optical and radio identifications. Some of these sources have been
identified in the RIXOS program.
}
        \maketitle

\section{Introduction}
     \n6217 is a B$_{\rm T}$ = 11.86 mag barred spiral, RSBbc(s)II, as
classified in the Revised-Shapley Ames Catalog of Bright Galaxies 
\citep{1981rsac.book.....S}. 
It was first designated as abnormal in the Atlas of Peculiar
Galaxies \citep{1966ApJS...14....1A,1966apg..book.....A}. 
In that classification it was No. 185, narrow filaments - 
between Nos. 149-152, jets and Nos. 209-215, material ejected from nuclei.  
The feature responsible for this classification was a narrow spiral arm like 
filament coming off the end of the inner bar, much less wound up than the 
inner arms, which gave the impression of ejected material.

Subsequently the galaxy was described as ``nuclear star burst, UV flat spectrum,
infra red bright" Seyfert 3 \citep{1998A&A...334..439B,1998csan.book.....V}.
Observations of optical
spectra dominated by stellar photoionization \citep{1992ApJ...388..310K}
and of extended radio emission
at 1.46 and 4.9 GHz \citep{1984A&A...134..207H} and CO emission strongly peaked
towards the center of the galaxy \citep{1996A&AS..115..439E} support this
classification. In a Hubble Space Telescope imaging survey of nearby
active galactic nuclei \citet{1998ApJS..117...25M} classify the nuclear
region (inner few arcsec) as CL, i.e. cluster, lumpy \ion{H}{ii} region, knots,
and DC, i.e. dust disk/dust lane passing close or through center.

\n6217 was detected in X-rays with the satellites Ginga and ROSAT (2--10 keV
and 0.1--2.4 keV, respectively)
as discussed in a multi-wavelength catalog of Seyfert 2 galaxies observed in 
the 2--10 keV energy band by \citet{1996ApJS..106..399P}. In addition, 
the galaxy is  identified in one of the fields selected for the ROSAT 
International X-ray/Optical
Survey \citep[RIXOS,][]{2000MNRAS.311..456M} as source number 122-16. 
The narrow emission line galaxy
(NELG) has been classified as a weak [\ion{O}{i}] LINER with a 
power law photon index of the ROSAT spectrum of $1.7\pm0.1$ and time 
variability below 14\%  \citep{1997MNRAS.285..831N,1999MNRAS.308..233M}.

In this paper we will report on a careful analysis of all ROSAT observations
of the \n6217 field aiming for the highest sensitivity well beyond the
RIXOS threshold of 3\ergcm{-14} (0.5--2.0 keV) for point sources
and diffuse emission. This led to the detection of a possible one-sided X-ray jet
from \n6217 and a serendipitous PSPC survey at a Galactic viewing angle with
moderate foreground absorption (L$_{\rm II}$ = 111\fdg2, 
B$_{\rm II}$ = 33\fdg5, \nh = 4.1\hcm{20}
\citep{1990ARA&A..28..215D})
nearly as deep as dedicated ROSAT PSPC surveys 
\citep[e.g.][]{1998A&A...329..482H}. 
Throughout the paper we will assume
a distance to \n6217 of 24.6 Mpc, i.e. 1$'\cor7.2$~kpc 
\citep{1990ApJS...73..359C}.

\section{Observations and results}
\begin{figure}
  \resizebox{\hsize}{!}{\includegraphics[bb=54 254 508 673,clip]{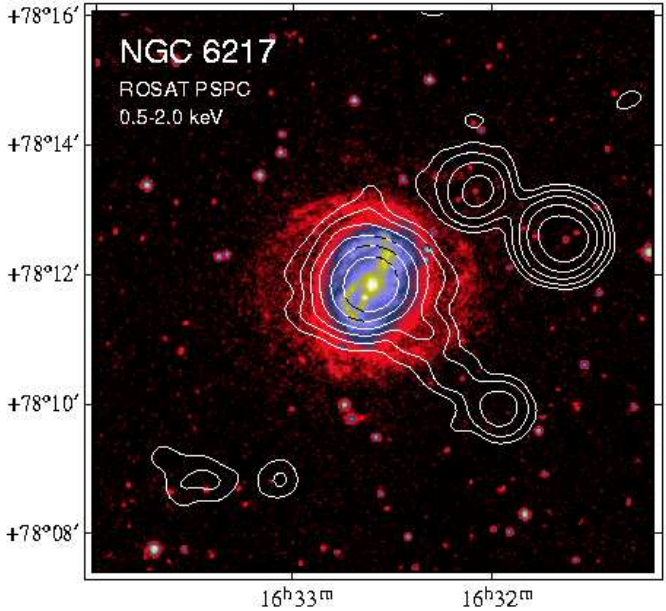}}
    \caption[]{Contour plot of the field around \n6217 for ROSAT PSPC hard band
    (0.5--2.0 keV) superimposed on an optical image extracted from the digitized
    sky survey II blue. The X-ray image has been smoothed according to the 
    on-axis point spread function, X-ray contours are given in units of
    $\sigma$ ($60 \cdot 10^{-6}$ ct s$^{-1}$ arcmin$^{-2}$) above the
    background ($170 \cdot 10^{-6}$ ct s$^{-1}$ arcmin$^{-2}$). Contour levels
    are 5, 10, 20 , 40, 80, and 160 units
    }
     \label{xopt}
\end{figure}

\begin{figure}
  \resizebox{\hsize}{!}{\includegraphics[angle=-90,clip]{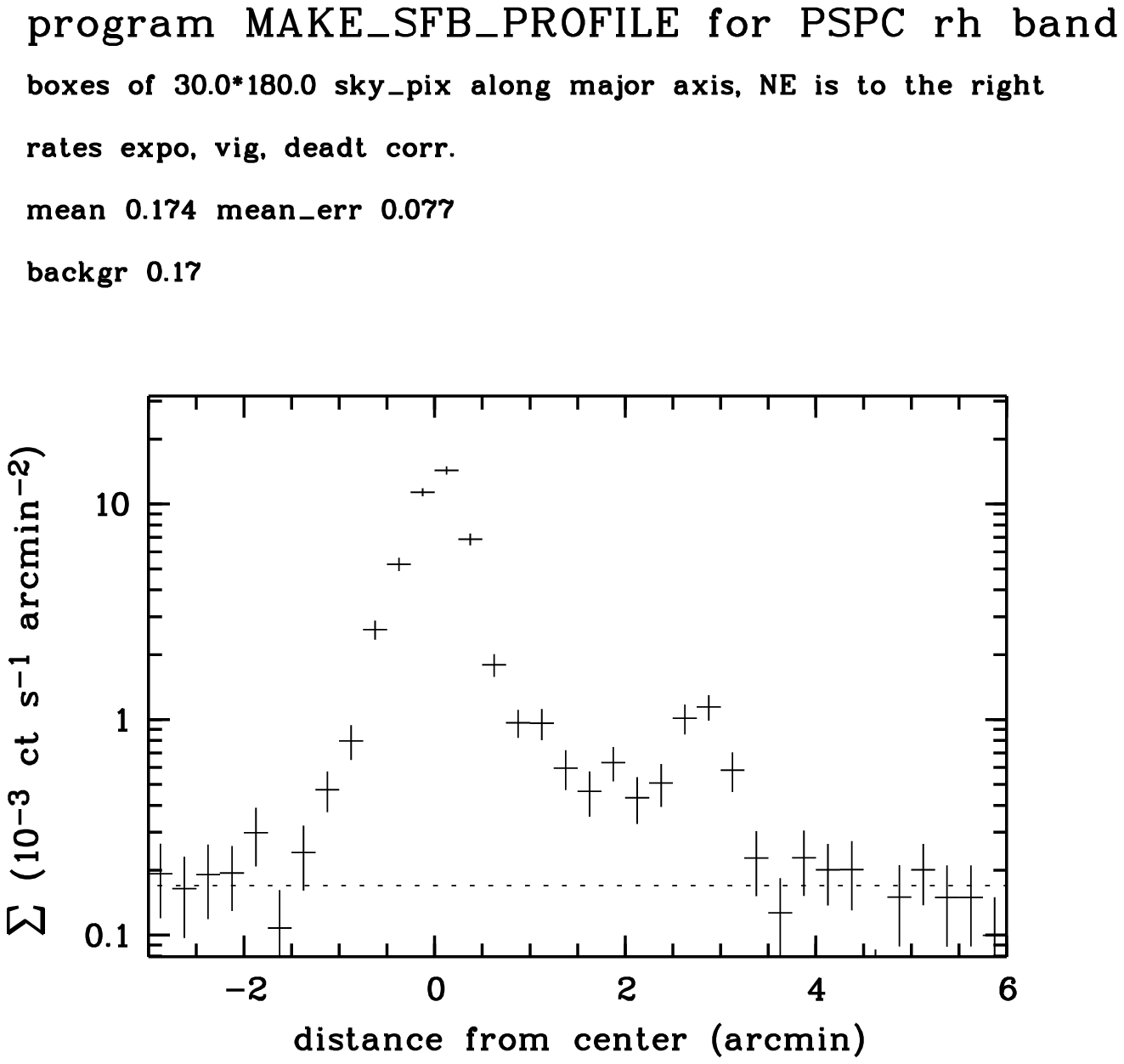}}
    \caption[]{
     Spatial distribution of surface brightness along position angle 226\degr\ 
     versus the distance from the \n6217 center.
     ROSAT PSPC hard (0.5--2.0 keV) counts are integrated in boxes of
     $15\arcsec\times 90\arcsec$. 
     The dotted line indicates the background
     surface brightness determined at distances of $\ge1\farcm5$ NE and 
     $\ge3\farcm5$ SW of \n6217}
     \label{sfb}
\end{figure}

\begin{figure}
  \resizebox{\hsize}{!}{\includegraphics[bb=54 254 508 673,clip]{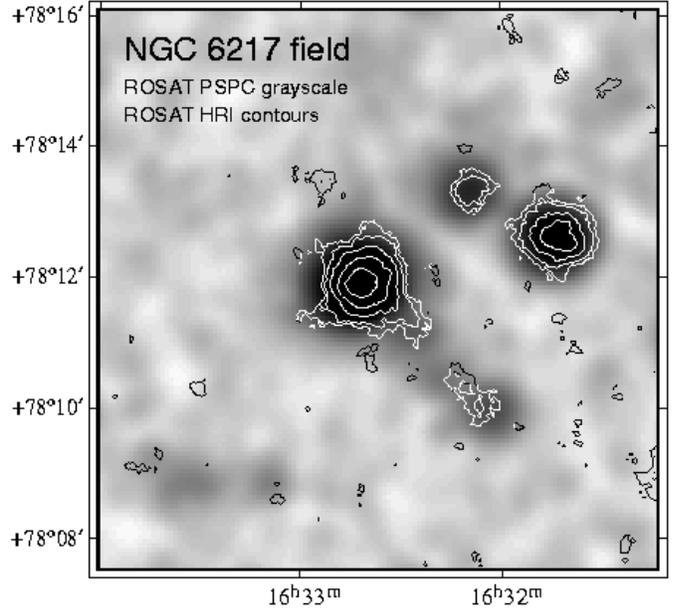}}
    \caption[]{Contour plot of the field around \n6217 from an adaptively smoothed
    ROSAT HRI image (see text) superimposed on a ROSAT PSPC hard band (0.5--2.0 keV)
    image. The PSPC image has been smoothed according to the 
    on-axis point spread function. HRI X-ray contours are given in units of
    $\sigma$ ($215 \cdot 10^{-6}$ ct s$^{-1}$ arcmin$^{-2}$) above the
    background ($2100 \cdot 10^{-6}$ ct s$^{-1}$ arcmin$^{-2}$). Contour levels
    are 3, 5, 10, 20 , and 40 units
    }
     \label{hripspc}
\end{figure}

\begin{figure}
  \resizebox{\hsize}{!}{\includegraphics[bb=80 54 385 395,angle=-90,clip]{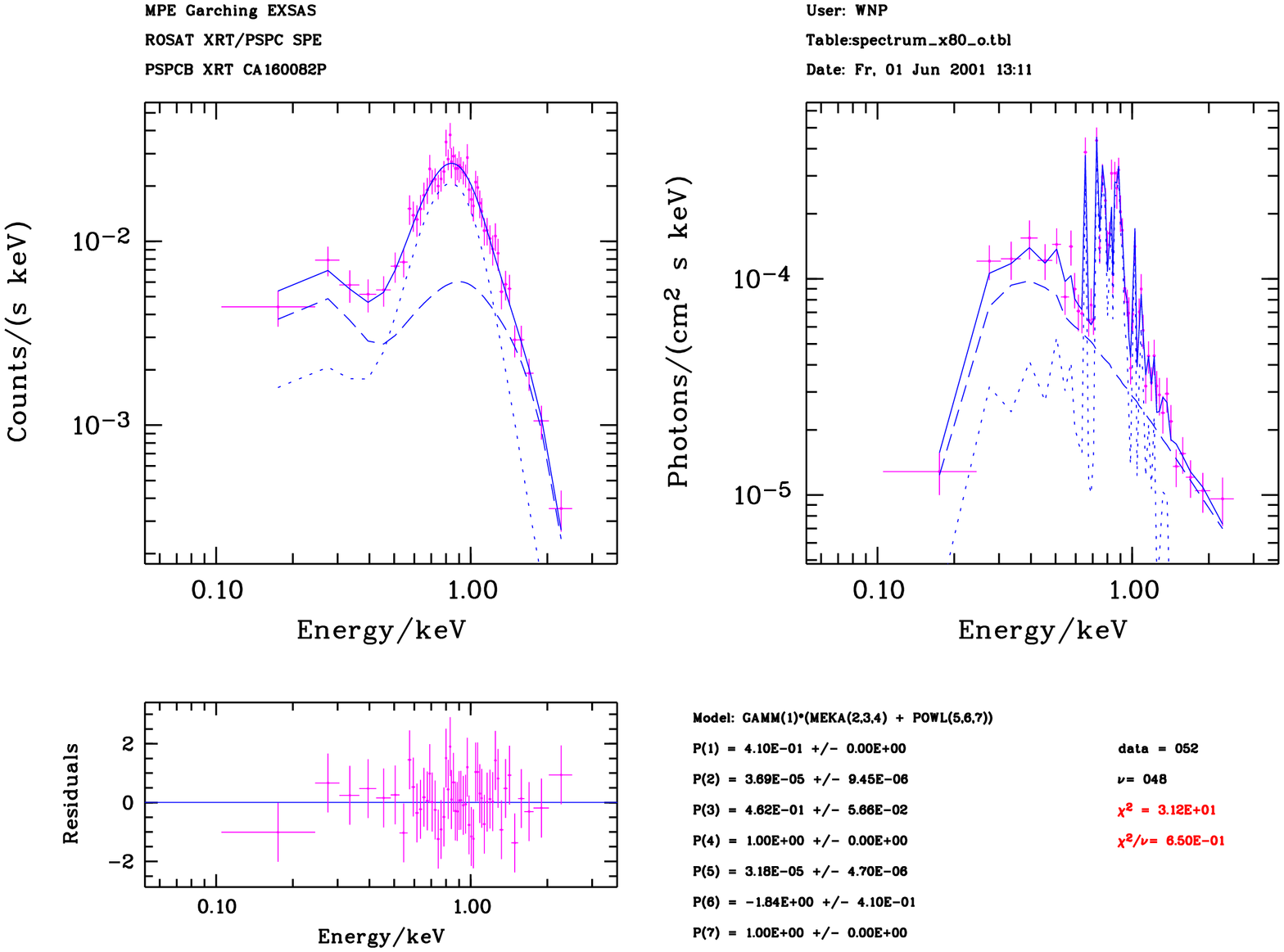}}
    \caption[]{Two component fit to the emission of \n6217. Count rate at the detector
    normalized to the energy, in counts s$^{-1}$ cm$^{-2}$, the crosses represent the
    measured count rates, the solid curve gives the best fit, the dotted curve represents
    the thin thermal, the dashed to the power law component (cf. Sect. 2)
    }
     \label{spectrum}
\end{figure}

During the ROSAT Wide Field Camera (WFC) `first light' observations 
the white-dwarf binary system \object{RE J1629+780} was serendipitously 
detected as a bright source \citep{1992Natur.355...61C}. \object{RE J1629+780} 
was then used throughout the entire ROSAT mission \citep{1982AdSpR...2..241T} 
as a WFC calibration source to monitor the gain and about a hundred PSPC and 
HRI \citep{1987SPIE..733..519P} exposures were taken in parallel to the WFC
observations. The total PSPC exposure 
time on \object{RE J1629+780} is 129\,930 s made up of 31 pointings. Observation 
142566p listed in the archive for the same pointing direction was rejected
due to an offset from this nominal pointing direction of $\sim$6\farcm4 mainly
in right ascension. We also did not include two short observations in February 
1998 to avoid detector degradation effects. The total HRI exposure time on 
\object{RE J1629+780} is 247\,247 s made 
up of 67 pointings. \n6217 is offset from \object{RE J1629+780} 
by $\sim 12$\arcmin\ to the NE and therefore during \object{RE J1629+780}
exposures well inside the ROSAT PSPC
inner support ring and also inside the HRI field.

Source detection and position determination were performed over the inner
$\sim 40\arcmin\ $field of view for each of the PSPC observations in the 
0.5--2.0 keV band \citep{1994EXSAS}. 
The event sets had to be individually shifted by up to 10\arcsec\ in right
ascension and 5\arcsec\ in declination with respect to an average  
position of the bright X-ray sources detected in each observation to correct 
residual boresight errors. The position
corrected data sets were then merged together. 
Though the PSPC observations were in general of rather low background, we
removed times of high background in the 0.5--2.0 keV band to increase the
sensitivity for faint structures which reduced the useful exposure time 
by $\sim 10\%$ to
117\,058 s. The source detection procedures were re-run and source positions
corrected using the RIXOS identifications \citep{2000MNRAS.311..456M} 
in the field reducing the systematic absolute position error to 0\farcs8
if one does not include an error due to the optically determined RIXOS 
positions. As \citet{2000MNRAS.311..456M} state that the positional accuracy in the 
RIXOS catalog is better than 1\arcsec, the residual systematic error will
only give a significant contribution to the position error of the brightest 
sources and is not corrected for in the 90\% uncertainty position error \perr. 
We also determined the likelihood of existence in the 0.5--2.0 keV band \exil\  
and for the best position hardness ratios defined as 
HR1 = (H--S)/(S+H) and HR2 = (H2--H1)/(H1+H2) where S, H, H1, and H2 
denote count rates in the 0.1--0.4 keV, 0.5--2.0 keV, 0.5--0.9 keV 
and 0.9--2.0 keV bands, respectively. Taking into account the Galactic 
foreground absorption and folding a power law spectrum with a photon
index of 2 through the instrument response we determine a energy to count rates 
conversion factor in ct s$^{-1}$ in the 0.5--2.0 keV band of 0.777 for a
source with a flux of \oergcm{-11}. Because the PSPC sensitivity is close to one
in the hard band and variable foreground absorption mainly influences the
soft band, this factor is almost insensitive to the assumed spectral shape
and a wide range of \nh.  

The resulting list of 91 sources detected with \exil$> 10$ 
in the $38\farcm3\times 38\farcm3$ field
centered on \object{RE J1629+780} is only a by-product from the analysis 
of the field around \n6217. It reaches, however, much deeper than the published 
RIXOS catalog entries of the field and we therefore included it in the appendix 
(Table~\ref{tab-cat}) together with an exposure corrected hard band X-ray image 
(Fig.~\ref{xcontours}) with numbered boxes marking the catalog entries. 
Table~\ref{tab-cat} gives ROSAT name (col. 1), source number (col. 2),
X-ray position (cols. 3--4), \perr (col. 5), \exil and count rate for the 
0.5--2.0 keV PSPC band (cols. 6--7), hardness ratios HR1 and HR2 (cols. 8--9), and 
some comments on possible identifications (col. 10). The faintest sources
in the catalog have a 0.5--2.0 keV flux of $\sim 2.2$\ergcm{-15}. 

The galaxy \n6217 is 13\farcm3 from the center of the field. There
seems to be an X-ray jet, emanating from the galaxy to about 2\farcm7 projected at 
a position angle of 226\degr$\pm$2\degr\ (Figs.~\ref{xcontours} and \ref{xopt}). The
X-ray surface brightness along the feature was integrated in 
$15\arcsec\times 90\arcsec$ boxes for the 0.5--2.0 keV PSPC band (Fig.~\ref{sfb}).
From the surface brightness distribution we determine background corrected count 
rates of (162, 8.5, 9.6)$\cdot 10^{-4}$\ct\ integrating the brightness at distances 
-1\farcm5--1\farcm0, 1\farcm0--2\farcm25, 2\farcm25--3\farcm5, respectively, which
can be attributed to \n6217, an intermediate region, and the SW `knot'. These
count rates are in good agreement with the source detection results (see
sources 73, 75, 78, and 80 of Table~\ref{tab-cat}) and correspond to 
(151, 7.9, 9.0)\ergs{38} at the distance of \n6217. Source 80 represents
\n6217, 73 the SW knot. Sources 75 and 78 are found in the bridge and at
the `tongue' extending from the galaxy. Though we chose a small extraction
radius for the source detection, specifically the results derived for 
source 78 and to a lesser amount for 75 may suffer from cross talk from
sources 80 and 73. 

The count rate detected for \n6217 is within the errors
the same as reported in the RIXOS survey \citep{2000MNRAS.311..456M} 
supporting the lack of time variability quoted by \citet{1997MNRAS.285..831N}.
\citet{1999MNRAS.308..233M} characterized the X-ray spectra of the
extragalactic objects in the RIXOS survey based on power law fits on three broad
band X-ray colors assuming galactic absorption. This methode of cause only gives
reasonable fits if the intrinsic spectrum really can be described by a power
law. The number of photons of the RIXOS AGN identifications (source 39, 60, 68,
and 88) as well as in \n6217 were sufficient to do fits with more than 10 and up 
to 52 data points in \n6217. 
For the AGN sample power law models gave acceptable 
fits with parameters coinciding within the errors with the RIXOS results. For
\n6217 however, simple power law (and also thermal bremsstrahlung or thin
thermal) models did not give acceptable fits (reduced $\chi^2 > 1.8$) even if 
the absorption column was not fixed to galactic. A combination of a thin thermal 
plasma of solar abundance and a power law with absorption for both components 
fixed to the galactic foreground resulted in an acceptable fit 
(reduced $\chi^2 = 0.66$). The temperature of the thermal component was 
$0.51\pm0.06$ keV and the power law photon index $1.8\pm0.3$ with absorption
corrected luminosities in the 0.1--2.4 keV band of (1.1 and 1.0)\ergs{40} (
Fig.~\ref{spectrum}).

There are not enough counts in the jet-like feature to do a detailed spectral
analysis. However, the hardness ratios in Table~\ref{tab-cat} of the sources 
detected therein indicate a somewhat harder spectrum than the power law
spectrum with photon index of 1.7 of \n6217 \citep[see][and HR1 1.0, HR2
$\sim$ 0.2 compared to 0.89 and 0.00]{1999MNRAS.308..233M}. While the feature
is clearly extended from \n6217 to the SW it can not be resolved by the
PSPC in the perpendicular direction. As Fig.~\ref{sfb} demonstrates, there is 
no indication of a similar feature on the opposite side of the galaxy. 

To analyze the HRI observations we merged the event sets, corrected the pointing
directions with a time resolution of 40 s using the bright soft source
\object{RE J1629+780} as reference. Finally, we screened the data for times of 
high background which reduced the useful time of exposure by more than 27\% to 
179\,793 s. Because the PSPC collects at least 2.5 times the photons 
and in addition has a lower background than the HRI the faint emission to the
SW of \n6217 is just marginally detected. Due to the relatively large off-axis
position of \n6217 one also does not gain much due to the better intrinsic spatial 
resolution of the HRI as the point spread function is dominated by the X-ray
telescope \citep{1988ApOpt..27.1404A}. However, the HRI results still can
be used to confirm the PSPC result. 
To further reduce the background due to UV emission and cosmic rays, we 
used HRI raw channels 2--8. An HRI image (0.1--2.4 keV) was constructed 
with a bin size of 2\farcs5 and adaptively smoothed by convolution with a circular 
top hat kernel using 100 counts. Iso-intensity contours of the resulting HRI image
are superimposed on a PSPC 0.5--2.0 keV gray-scale image in Fig. ~\ref{hripspc} and 
nicely confirm the brighter PSPC structures with slightly better resolution. 

\section{Discussion}
\begin{figure}
  \resizebox{\hsize}{!}{\includegraphics[bb=54 254 508 673,clip]{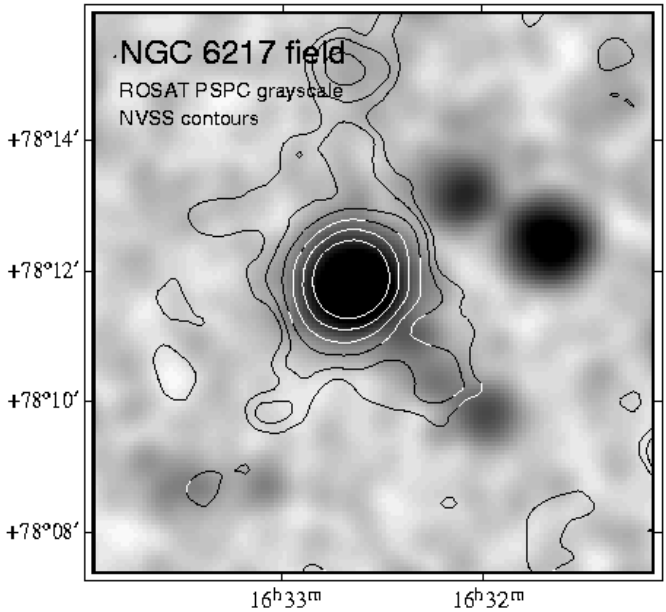}}
    \caption[]{Contour plot of the field around \n6217 from the NVSS Stokes I
    image (restoring beam width 45\arcsec\ FWHM, rms brightness fluctuations
    about 0.45 mJy/beam) superimposed on the ROSAT PSPC 
    hard band (0.5--2.0 keV) image from Fig.~\ref{hripspc}. Radio contours are 
    1, 2, 4, 8, 16, and 32 mJy/beam}
     \label{nvsspspc}
\end{figure}

\begin{table*}
\caption[]{Parameters of X-ray jets.}
\begin{tabular}{lrrrrll}
\hline\noalign{\smallskip}
\multicolumn{1}{l}{Source}  & \multicolumn{1}{l}{Distance} &
\multicolumn{2}{l}{Extent} & 
\multicolumn{1}{l}{$L_{\rm X}$}    & \multicolumn{1}{l}{Comments}  & \multicolumn{1}{l}{Ref.}\\ 
\noalign{\smallskip}
& \multicolumn{1}{c}{[Mpc]}    & \multicolumn{1}{c}{[\arcsec]} &
\multicolumn{1}{c}{[kpc]} & \multicolumn{1}{c}{[erg s$^{-1}$]} & & \\
\noalign{\smallskip}\hline\noalign{\smallskip}
\object{Cen A}        &  3.5 &  240 &  2.8 & $2.4\cdot10^{39}$ & 0.1--2.4 keV band, radio, faint X-ray counter-jet &  1,2\\
\object{M87}          & 17.0 & 11.5 &  1.0 & $1.9\cdot10^{40}$ & 2--10 keV band, radio, optical, one-sided        &  3  \\
\n6217                & 24.6 &  160 & 18.8 & $1.7\cdot10^{39}$ & 0.5--2.0 keV band, radio, one-sided              &   this work  \\
\object{Pictor A}     &  151 &  115 &   85 & $6.8\cdot10^{40}$ & 2--10 keV band, radio, one-sided, hot spot            &  4  \\
\object{3C273}        &  550 &   22 &   60 & $4.0\cdot10^{43}$ & 0.5--8 keV band, radio, optical, one-sided            &  5  \\
\object{PKS 0637-752} & 1200 &   12 &   70 & $1.5\cdot10^{44}$ & 2--10 keV band, radio, optical, one-sided             &  6  \\
\noalign{\smallskip}
\hline
\noalign{\smallskip}
\end{tabular}
\label{jets}

References and notes:\\
(1) ROSAT HRI, \citet{1996ApJ...470L..15D};
(2) Chandra, \citet{2000ApJ...531L...9K};
(3) XMM-Newton, \citet{2001A&A...365L.181B};
(4) Chandra, $L_{\rm X}$ of hot spot $8.4\cdot10^{41}$ erg s$^{-1}$, \citet{2001ApJ...547..740W};
(5) Chandra, \citet{2001ApJ...549L.167M};
(6) Chandra, \citet{2000ApJ...540L..69S}
\end{table*}

Both, in the PSPC and the HRI the main part (at least $\sim$80\%) of the X-ray emission of 
\n6217 is not resolved and appears as a point-like source with a jet-like feature emanating
to the SW. The thin thermal component of the unabsorbed X-ray spectrum of 
\n6217 most likely arises in the nuclear starburst area of the galaxy, 
while the power law may reflect the unabsorbed jet component
within a few kpc of the nucleus. The active nucleus most likely is heavily obscured and 
-- similar to many other Seyfert 2 nuclei -- not detected in the ROSAT band 
\citep[see e.g.][]{moran2001}. The radio flux density of \n6217 is close to that of the
optical flux density and the galaxy therefore has to be 
classified as radio quiet \citep{1989AJ.....98.1195K}.

Before discussing the X-ray feature to the SW of \n6217 as an X-ray jet we want 
to investigate the question as to whether the extent could be an accidental alignment of
unrelated X-ray sources. Figs.~\ref{xopt}, \ref{xcontours}, and also Fig.~\ref{hripspc} 
show that this could be possible. However, it would require at least three very
closely spaced sources to be aligned rather exactly with the nucleus of the 
galaxy.  This seems rather unlikely but certainly can not be excluded by the
PSPC and HRI observations. Therefore we have to look for additional arguments
from other wavelengths.

Radio observations provide additional evidence for jet-like emission from \n6217. 
There is not only arcsec-scale extended emission from the nuclear area
\citep{1984A&A...134..207H}, but also VLA radio maps at 1.49 GHz 
\citep{1987ApJS...65..485C} 
and NRAO VLA Sky Survey \citep[NVSS,][]{1998AJ....115.1693C}
reveal a complicated extended radio source on arcmin-scale with an extension 
coming out to $\sim$2\arcmin, exactly along the 
line of the X-ray extension (see Fig.~\ref{nvsspspc}) which might represent
a radio jet. Unfortunately, the resolution and sensitivity of the observations
is not sufficient to make a clear radio jet identification. 
     
There is no optical correspondence at the position of the X-ray feature 
on either the POSS II plates  
(see Fig.~\ref{xopt} for an overlay of PSPC 0.5--2.0 keV band contours on 
the digitized POSS II blue image) or the 200-inch telescope photograph. 
Also the Hubble Space Telescope imaging survey of nearby
active galactic nuclei \citep{1998ApJS..117...25M} reveals no clear hint of a jet
from the inner few arcsec around the nucleus. Since the interior of the
galaxy is barred, it is difficult to judge the position of the minor axis. But 
using the outer isophotes of the Schmidt plates it is possible to estimate that 
the direction of the x-ray feature is in the quadrant in which the minor axis 
lies (and also perpendicular to the bar). On the other hand, there are also no
clear optical counterparts along the X-ray feature as might be expected if 
it were an accidental alignment of unrelated sources. Such
counterparts have been found for many of the other sources of similar
X-ray brightness in the field (see Table~\ref{tab-cat}).

For the further discussion we assume that the X-ray feature originates
from the nuclear area of \n6217. This then would point at two possible 
explanations for the feature, a galactic superwind driven by a nuclear
starburst or a one-sided jet originating from an active nucleus in \n6217. 

Nuclear starbursts have been observed for several nearby galaxies leading 
to X-ray emission far out in the halos of galaxies 
\citep[e.g.][]{1998ApJS..118..401D}. However,there the X-ray emission only stays 
collimated within the disk of the galaxies and fills large parts of the
galaxy halos, as e.g. the plume and outer halo emission in \object{NGC 253} 
\citep{2000A&A...360...24P} or the nuclear superbubble and halo emission of 
\object{NGC 3079} \citep{1998A&A...340..351P} show. 
Also the spectrum of the emission in the halo in general is significantly softer 
than the emission from the galaxy nuclear area and disk. No knot-like radio
emission has been reported from X-ray halos fed by starbursts. Therefore a 
galactic superwind is a rather unlikely explanation.

This leaves the interpretation as a one-sided X-ray jet. 
The X-ray morphology would indicate that at least three knots with distances of
$\sim$1\arcmin, 1\farcm8, and 2\farcm7 from the nucleus contribute to the emission. 
The knot at 1\farcm8 also shows up as a radio knot (see Fig.~\ref{nvsspspc}). 

X-ray jets with and without knot-like substructure have been reported for several AGN.
The most famous X-ray jets are those in \object{Cen A}, \object{M87}, 
\object{Pictor A}, \object{3C273}, and \object{PKS 0637-752}, all radio-loud objects. 
However there is also the precessing jet in  the radio-quiet galaxy
\object{NGC 4258} \citep{1994A&A...284..386P,2000ApJ...536..675C} that
also manifests itself in X-rays in the anomalous arm structure seen in 
projection onto the disk in the halo of this LINER/Seyfert galaxy. 
X-ray images from Chandra are now resolving the spatial structure along 
quasar jets producing remarkable images. In Table~\ref{jets} we have
put together properties of these jets collected mainly from the new
Chandra observations to compare with the \n6217 jet properties. To 
allow easier comparison, we have converted all numbers assuming
$q_0 =0.5$ and $H_0 = 70$ km s$^{-1}$ Mpc$^{-1}$. We give distance, apparent
projected X-ray dimension in arcsec and kpc, respectively,  the X-ray luminosity
and we comment on the
energy band used for the luminosity determination, and on radio and optical
identification of the jets and give references.
These X-ray jets are all strongly one-sided
as would be the proposed jet in \n6217 and are all also observed as radio 
jets or as colinear radio lobes. They also show hard X-ray spectra as
proposed from the hardness ratios for the \n6217 jet.

According to Table~\ref{jets} the \n6217 jet would be one of the longest in apparent dimensions and
would seem to merit higher resolution from Chandra or greater wavelength 
coverage investigations from XMM-Newton in order to learn more about the 
characteristics of energetic ejections. Its measured luminosity is only
moderate. However, jet emission may in part be hidden within the flux attributed to \n6217
itself becaused of the limited resolution of the ROSAT PSPC, as already is indicated by
the slightly better resolution of the HRI which detects jet emission closer to
the galaxies nucleus (c.f. Fig.~\ref{hripspc}). The missing optical evidence for
a jet in \n6217 is no strong argument. 
Also in \object{Cen A} there is no particular optical 
object associated with the X-ray jet although there are some outer gaseous 
filaments which lie along the projected track and in \object{Pictor A} 
there is no 
optical jet visible even in large telescope images. While this together with
the faintness in X-ray and radio may explain
that we do not see hints for jet emission in the optical, deeper, large 
telescope imaging and radio observations would strongly be needed in addition
to better X-ray data to put more strength to the proposed jet interpretation.

\begin{acknowledgements}
The X-ray data used in this work were all obtained from the ROSAT Data
Archive at the Max-Planck-Institut f\"ur extraterrestrische Physik (MPE) at 
Garching. This research has made use of the NASA/IPAC Extragalactic Database 
(NED) which is operated by the Jet Propulsion Laboratory, California
Institute of Technology, under contract with the National Aeronautics and 
Space Administration.
To overlay the X-ray data we used an blue digitized sky survey II image.
The compressed files of the "Palomar Observatory - Space Telescope
Science Institute Digital Sky Survey" of the northern sky, based on
scans of the Second Palomar Sky Survey are copyright (c) 1993-1997
by the California Institute of Technology. All material not subject to the 
above copyright provision is  copyright (c) 1997 by the Association of 
Universities for Research in Astronomy, Inc. Produced under Contract 
No. NAS5-2555 with the National Aeronautics and Space Administration.
The ROSAT project is supported by the German Bundesministerium
f\"ur Bildung, Wissenschaft, Forschung und Technologie (BMBF/DLR)
and by the Max-Planck-Gesellschaft (MPG).
\end{acknowledgements}
\bibliographystyle{apj}
\bibliography{./MS1398}

\begin{appendix}
\section{Identification of X-ray sources detected in the field}
\begin{figure*}
  \resizebox{12cm}{!}{\includegraphics[bb= 53 255 506 672,width=12cm,clip=]{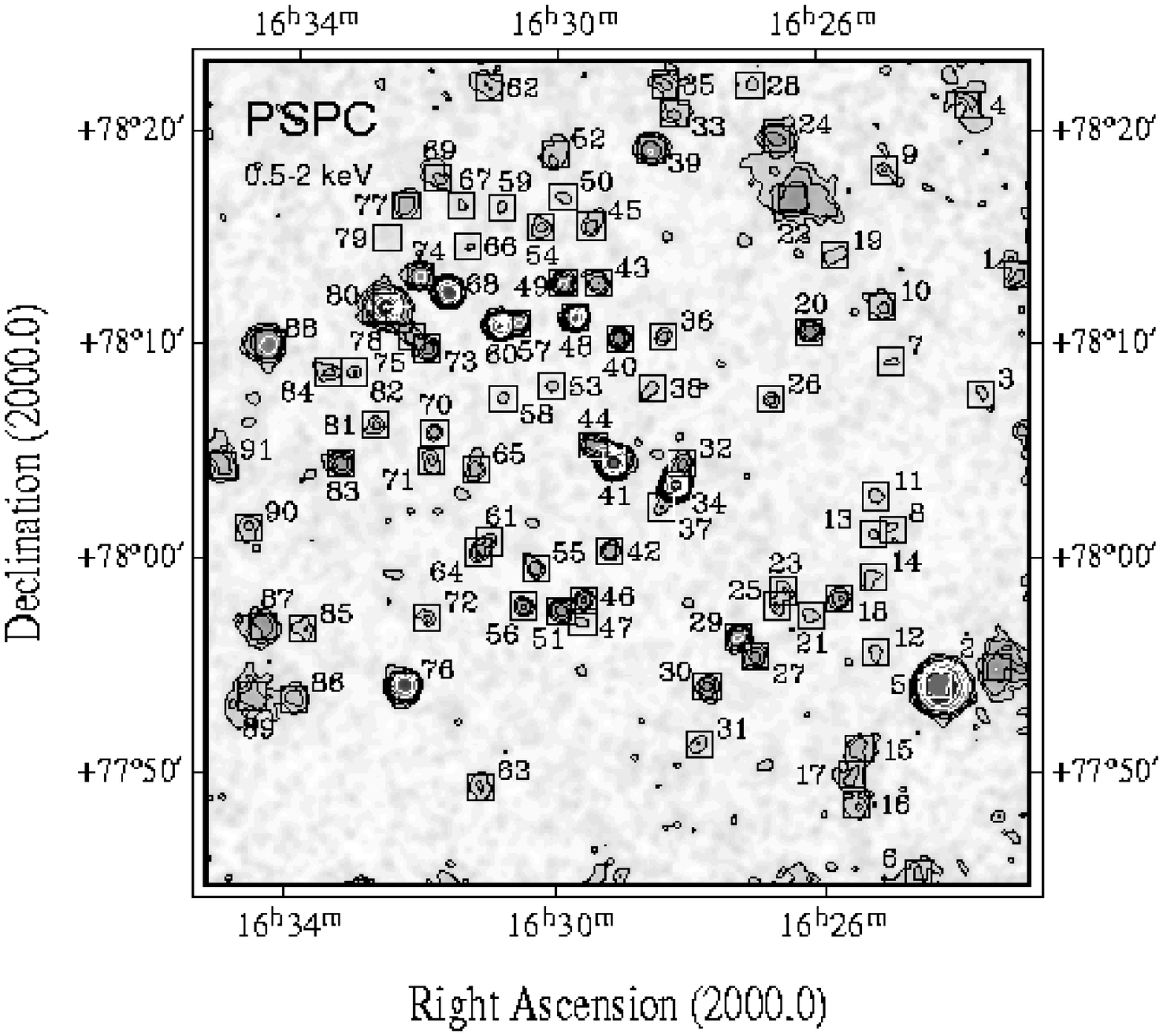}}
  \hfill
  \parbox[b]{55mm}{
    \caption[]{Contour plot superimposed on exposure corrected and smoothed 
    gray-scale image of the ROSAT PSPC hard band (0.5-2 keV, 117 ks low
    background integration time) of the $38\farcm3\times 38\farcm3$ field
    centered on \object{RE J1629+780}. Contour levels are the same as in 
    Fig.~\ref{xopt}.
    Sources from the catalog are given as boxes with source numbers   
     }
    \label{xcontours}}
\end{figure*}
Ninety-one sources are detected in the $38\farcm3\times 38\farcm3$ field
centered on \object{RE J1629+780} in the deep PSPC exposure. Table~\ref{tab-cat} shows
their characteristics. Sources 5, 22, 34, 41, 44, 57, 68, 78, and 80 have been 
flagged as extended by the detection algorithm. While for most of them
this might indicate overlapping PSF of nearby point sources, extended emission 
in 22 further confirms the galaxy cluster identification of the RIXOS program.
Extended emission in 80 reflects overlapping emission components from the
hot interstellar medium and bright point sources in the \n6217. The cases
of source 5 and 41/44 will be discussed separately.

To guide optical identifications we produced overlay plots 
of PSPC 0.5-2.0 keV band contours on the DSS2 red and blue plates and inspected
APM \citep{1994Spec....2...14I} finding charts and the USNO-A2 Catalogue 
\citep{1998USNO2.C......0M} from which we get positional accuracies for the
optical candidates of better than 1\arcsec\ for USNO-A2 (2\arcsec\ for APM), 
optical magnitudes of the blue (O) and red (E) plates, and the color index O-E. 
Optical counterparts were searched for each in the $\sim$\perr\ X-ray X-ray error 
circle. If at all usually only one optical source correlates. 
For 33 sources we give the higher accuracy USNO-A2 identification 
candidate information in col. 10
of Table~\ref{tab-cat} (source number with magnitude from blue plate, 
the color index, and separation in arcsec between optical and X-ray position 
following in parenthesis). For source 42 only APM information is available. 
For sources 54, 56, and 62 two USNO-A2 sources are candidates for the
identification. 

The bright source 5 includes 3 USNO-A2 sources within its extent. The
best correlation is given in the table. There are, however, two fainter
objects within 15\arcsec, i.e. U57709(18.2,0.7,12.6) and 
U57766(16.7,1.5,14.7). Comparing POSS I and II plates we find that the best 
correlating optical source U57802 is a newly detected fast moving object 
with a proper motion (-170 mas/yr in RA$\times$cos(Dec) and
110 mas/yr in Dec). Most certainly two or even all three optical
candidates contribute to the X-ray emission.
 
To investigate the reliability of the identifications and allow rough
classification in foreground stars and AGN, 
we have added in col. 11 log(f$_{\rm x}$/f$_{\rm v}$) = log(f$_{\rm x}$) + 
(m$_{\rm v}$/2.5) + 5.37 \citep[see][]{1988ApJ...326..680M}.
This X-ray to optical luminosity relation can be used
to discriminate between stars, for which log(f$_{\rm x}$/f$_{\rm v}$)
is generally $<$ -1, and  AGN for which this quantity is in the range  -1.2 
to +1.2 \citep[see also][]{1998A&A...340..351P,1999A&A...342..101V}.
X-ray fluxes in the 0.5--2.0 keV band have been
determined from the count rates assuming a power law spectrum with photon 
index of 2, corrected for Galactic absorption (see Sect. 2).  For 
f$_{\rm v}$ we used the red magnitude of the first identification 
candidate. 

Some of the brighter X-ray sources in Fig.~\ref{xcontours} (sources 22,
39, 48, 68, 76, 80, and 88) have already been investigated in the RIXOS program 
\citep[][see col. 10 of Table~\ref{tab-cat}]{1997MNRAS.285..831N,1997MNRAS.291..177P,2000MNRAS.311..456M}. 
They were identified with the help of optical spectroscopy as a cluster of
galaxies, AGN, star, AGN, AGN, star, emission line galaxy, and AGN,
respectively. The AGN and stars nicely confirm the  
log(f$_{\rm x}$/f$_{\rm v}$) classification scheme.
One of the AGN (source 60 with a redshift of 0.358) is very red and could be 
absorbed by some outlying dust from \n6217. This would be a further incentive 
to spectroscopically identify the  fainter candidates near the galaxy.

While in the soft band the emission of \object{RE J1629+780} dominates the
central ares of the field, in the 0.5--2.0 keV band there is a smooth transition
from source 41 (i.e. \object{RE J1629+780}) to 44 indicating the contribution
from even another source at intermediate position which from the optical
overlays can be identified with a blue stellar object cataloged in APM
with a blue magnitude of 20.1, color 0.1, at position (epoch J2000.0)
RA = $16^{\rm h}29 ^{\rm m}$18\fs9, Dec = +78\degr05\arcmin14\arcsec. 

We find five correlations of the X-ray source list with the 
1.4 GHz NRAO/VLA Sky Survey 
\citep[NVSS,][]{1998AJ....115.1693C} radio source catalogs, one of which is also  
contained in the 325 MHz 
Westerbork Northern Sky Survey \citep[WENSS,][]{1997A&AS..124..259R}
catalog. The brightest source is \n6217. The second brightest source is also
identified with a optical source and most likely is, based on the optical 
color and log(f$_{\rm x}$/f$_{\rm v}$), a radio galaxy.
The other three sources have no optical counterparts and are most likely
radio-loud AGN.
 
\end{appendix}

\clearpage
\onecolumn
\landscape
\setcounter{table}{1}
\begin{table}
\caption[]{X-ray sources detected with the ROSAT PSPC in the 0.5-2 keV band 
           in a field of 38\farcm3$\times$38\farcm3 centered on 
	   \object{RE J1629+780}. }
\scriptsize
\begin{tabular}{rrrrrrrrrp{70mm}l}
\hline\noalign{\smallskip}
\multicolumn{1}{c}{1} & \multicolumn{1}{c}{2} & \multicolumn{1}{c}{3} &
\multicolumn{1}{c}{4} & \multicolumn{1}{c}{5} & \multicolumn{1}{c}{6} &
\multicolumn{1}{c}{7} & \multicolumn{1}{c}{8} & \multicolumn{1}{c}{9} &
\multicolumn{1}{l}{10}  & \multicolumn{1}{c}{11}\\
\hline\noalign{\smallskip}
\multicolumn{1}{c}{Source name} &\multicolumn{1}{c}{No} & \multicolumn{2}{c}{RA~~~(J2000.0)~~~Dec} &
\multicolumn{1}{c}{\perr} & \multicolumn{1}{c}{\exil} &
\multicolumn{1}{c}{Count rate} & \multicolumn{1}{c}{HR1} & \multicolumn{1}{c}{HR2} & Identifications & 
\multicolumn{1}{c}{log(${\rm f}_{\rm x} \over {\rm f}_{\rm v}$)}\\
 & & \multicolumn{1}{c}{[h m s]} & \multicolumn{1}{c}{[\degr\ \arcmin\ \arcsec]} &
\multicolumn{1}{c}{[\arcsec]} & & \multicolumn{1}{c}{[\ct]} & & & &\\
\noalign{\smallskip}\hline\noalign{\smallskip}
RX J162300.6+781315 &   1 & 16 23 00.64 &78 13 16.0 & 19.5 &    20.8 & 5.08e-04 $\pm$1.0e-04 &  0.31$\pm$0.25 & -0.31$\pm$0.19 & U56154(17.0,0.5,23.9)                                        & -2.2 \\ 
RX J162327.7+775453 &   2 & 16 23 27.78 &77 54 53.2 &  6.6 &   266.1 & 2.17e-03 $\pm$1.7e-04 &  0.24$\pm$0.07 &  0.03$\pm$0.08 &                                                              &      \\ 
RX J162335.8+780740 &   3 & 16 23 35.89 &78 07 40.0 & 17.3 &    12.7 & 2.54e-04 $\pm$6.9e-05 &  1.00$\pm$0.00 &  0.20$\pm$0.26 & U56725(19.3,-0.1,14.9)                                       & -1.4 \\ 
RX J162341.7+782114 &   4 & 16 23 41.79 &78 21 14.2 & 21.7 &    19.0 & 6.96e-04 $\pm$1.4e-04 &  1.00$\pm$0.00 &  1.00$\pm$0.00 & U56892(16.0,0.6,10.5)                                        & -2.5 \\ 
RX J162420.6+775408 &   5 & 16 24 20.63 &77 54 08.4 &  1.1 & 14673.9 & 4.31e-02 $\pm$6.7e-04 &  0.17$\pm$0.01 &  0.06$\pm$0.02 & U57802(14.8,1.9,4.3)                                         & -1.7 \\ 
RX J162441.2+774520 &   6 & 16 24 41.29 &77 45 20.1 & 20.2 &    17.3 & 6.73e-04 $\pm$1.4e-04 &  1.00$\pm$0.00 &  0.44$\pm$0.17 &                                                              &      \\ 
RX J162457.6+780918 &   7 & 16 24 57.66 &78 09 18.2 & 15.9 &    12.0 & 2.53e-04 $\pm$6.9e-05 &  1.00$\pm$0.00 & -0.09$\pm$0.31 & U58672(16.6,0.6,14.1)                                        & -2.7 \\ 
RX J162459.0+780127 &   8 & 16 24 59.05 &78 01 27.8 & 16.2 &    14.2 & 2.99e-04 $\pm$7.4e-05 &  1.00$\pm$0.00 &  0.17$\pm$0.25 &                                                              &      \\ 
RX J162500.2+781814 &   9 & 16 25 00.29 &78 18 15.0 & 17.8 &    14.3 & 3.35e-04 $\pm$8.4e-05 &  1.00$\pm$0.00 &  0.14$\pm$0.27 &                                                              &      \\ 
RX J162504.4+781152 &  10 & 16 25 04.43 &78 11 52.6 & 10.0 &    46.2 & 6.37e-04 $\pm$9.8e-05 &  1.00$\pm$0.00 &  0.21$\pm$0.18 &                                                              &      \\ 
RX J162513.0+780300 &  11 & 16 25 13.02 &78 03 00.0 & 11.2 &    19.5 & 2.92e-04 $\pm$6.8e-05 &  1.00$\pm$0.00 &  0.16$\pm$0.22 & U58965(18.2,0.6,12.6)                                        & -2.0 \\ 
RX J162515.4+775542 &  12 & 16 25 15.49 &77 55 42.1 & 14.9 &    12.7 & 2.75e-04 $\pm$7.2e-05 &  1.00$\pm$0.00 &  0.33$\pm$0.25 & NVSS(9.2,1.9)                                                &      \\ 
RX J162515.6+780113 &  13 & 16 25 15.67 &78 01 13.6 & 13.6 &    13.6 & 2.37e-04 $\pm$6.4e-05 &  1.00$\pm$0.00 & -0.06$\pm$0.27 &                                                              &      \\ 
RX J162517.0+775914 &  14 & 16 25 17.10 &77 59 14.7 & 13.0 &    16.4 & 3.09e-04 $\pm$7.3e-05 &  1.00$\pm$0.00 &  0.73$\pm$0.22 & U59114(19.2,0.3,22.2)                                        & -1.5 \\ 
RX J162531.3+775111 &  15 & 16 25 31.32 &77 51 11.3 & 10.3 &    48.5 & 7.99e-04 $\pm$1.2e-04 &  0.97$\pm$0.27 &  0.23$\pm$0.12 & U59260(18.7,-0.3,8.8)                                        & -1.0 \\ 
RX J162534.4+774838 &  16 & 16 25 34.43 &77 48 38.5 & 18.8 &    13.8 & 3.98e-04 $\pm$9.6e-05 &  0.80$\pm$0.47 &  0.42$\pm$0.17 &                                                              &      \\ 
RX J162538.4+775002 &  17 & 16 25 38.46 &77 50 02.7 & 14.0 &    33.0 & 6.77e-04 $\pm$1.1e-04 &  1.00$\pm$0.00 &  0.35$\pm$0.12 & U59471(17.9,-0.2,6.5)                                        & -1.5 \\ 
RX J162547.3+775816 &  18 & 16 25 47.31 &77 58 16.1 &  6.6 &    64.3 & 6.47e-04 $\pm$9.1e-05 &  0.30$\pm$0.17 & -0.17$\pm$0.13 & U59661(13.8,0.8,5.7)                                         & -3.5 \\ 
RX J162547.6+781417 &  19 & 16 25 47.60 &78 14 17.7 & 14.4 &    16.8 & 3.55e-04 $\pm$8.2e-05 &  1.00$\pm$0.00 & -0.39$\pm$0.24 &                                                              &      \\ 
RX J162610.8+781045 &  20 & 16 26 10.87 &78 10 45.7 &  4.5 &   140.6 & 1.05e-03 $\pm$1.1e-04 &  1.00$\pm$0.00 &  0.16$\pm$0.10 &                                                              &      \\ 
RX J162613.2+775730 &  21 & 16 26 13.25 &77 57 30.2 & 13.8 &    10.5 & 2.25e-04 $\pm$6.5e-05 &  1.00$\pm$0.00 &  0.12$\pm$0.26 & U60332(16.9,0.4,10.1)                                        & -2.6 \\ 
RX J162624.7+781659 &  22 & 16 26 24.77 &78 16 59.7 &  4.6 &   327.7 & 2.86e-03 $\pm$1.9e-04 &  0.91$\pm$0.07 &  0.33$\pm$0.06 & F122\_552 R20.3 :GClustr :z=0.5                              &  0.1 \\ 
RX J162636.9+775838 &  23 & 16 26 36.99 &77 58 38.5 & 10.8 &    18.4 & 2.94e-04 $\pm$6.8e-05 &  1.00$\pm$0.00 &  0.03$\pm$0.21 &                                                              &      \\ 
RX J162639.6+781944 &  24 & 16 26 39.64 &78 19 44.2 &  6.2 &   193.3 & 1.91e-03 $\pm$1.6e-04 &  0.64$\pm$0.09 &  0.40$\pm$0.08 & U60881(17.9,-0.2,15.1)                                       & -1.0 \\ 
RX J162643.5+775757 &  25 & 16 26 43.52 &77 57 57.3 & 10.6 &    16.6 & 2.53e-04 $\pm$6.3e-05 &  1.00$\pm$0.00 &  0.65$\pm$0.21 &                                                              &      \\ 
RX J162646.7+780735 &  26 & 16 26 46.80 &78 07 35.9 &  7.8 &    24.3 & 3.00e-04 $\pm$6.4e-05 &  1.00$\pm$0.00 &  0.97$\pm$0.18 &                                                              &      \\ 
RX J162703.2+775534 &  27 & 16 27 03.27 &77 55 34.4 &  5.5 &   101.3 & 8.94e-04 $\pm$1.1e-04 &  0.70$\pm$0.16 &  0.04$\pm$0.11 &                                                              &      \\ 
RX J162703.7+782218 &  28 & 16 27 03.75 &78 22 18.4 & 21.8 &    10.1 & 2.60e-04 $\pm$7.6e-05 &  1.00$\pm$0.00 &  0.81$\pm$0.16 &                                                              &      \\ 
RX J162718.9+775628 &  29 & 16 27 18.91 &77 56 28.2 &  3.5 &   227.7 & 1.43e-03 $\pm$1.2e-04 &  0.67$\pm$0.10 &  0.37$\pm$0.08 & U61755(17.9,2.1,4.6), WN B1629.5+7802, NVSS(52.6,4.8)        & -2.0 \\ 
RX J162746.4+775411 &  30 & 16 27 46.45 &77 54 11.3 &  5.3 &    89.4 & 8.02e-04 $\pm$9.9e-05 &  1.00$\pm$0.00 &  0.38$\pm$0.11 &                                                              &      \\ 
RX J162754.0+775130 &  31 & 16 27 54.03 &77 51 30.5 & 12.3 &    15.5 & 3.00e-04 $\pm$7.3e-05 &  1.00$\pm$0.00 &  0.21$\pm$0.18 &                                                              &      \\ 
RX J162808.0+780437 &  32 & 16 28 08.05 &78 04 37.2 &  5.6 &    84.4 & 7.66e-04 $\pm$9.5e-05 & -0.20$\pm$0.09 &  0.18$\pm$0.10 & NVSS(4.8,3.2)                                                &      \\ 
RX J162813.3+782100 &  33 & 16 28 13.30 &78 21 00.2 & 10.3 &    57.7 & 7.94e-04 $\pm$1.1e-04 &  1.00$\pm$0.00 &  0.42$\pm$0.14 &                                                              &      \\ 
RX J162814.2+780336 &  34 & 16 28 14.25 &78 03 36.2 &  1.1 &  3308.8 & 9.51e-03 $\pm$3.0e-04 &  0.35$\pm$0.03 &  0.21$\pm$0.03 & U62860(17.6,-0.4,3.3), AGN z=0.64 (Y.Chu, priv.comm.)       & -0.3 \\ 
RX J162822.4+782221 &  35 & 16 28 22.46 &78 22 21.8 & 11.4 &    56.8 & 8.18e-04 $\pm$1.1e-04 &  1.00$\pm$0.00 &  0.59$\pm$0.12 &                                                              &      \\ 
RX J162824.8+781033 &  36 & 16 28 24.81 &78 10 33.9 &  7.7 &    26.9 & 3.12e-04 $\pm$6.5e-05 &  0.70$\pm$0.39 & -0.32$\pm$0.16 & U63068(14.8,1.5,7.9)                                         & -3.7 \\ 
RX J162826.5+780236 &  37 & 16 28 26.55 &78 02 36.4 & 12.6 &    13.7 & 2.07e-04 $\pm$5.6e-05 & -0.82$\pm$0.05 &  0.33$\pm$0.18 &                                                              &      \\ 
RX J162834.8+780810 &  38 & 16 28 34.84 &78 08 10.4 & 11.0 &    11.1 & 1.78e-04 $\pm$5.3e-05 &  1.00$\pm$0.00 & -0.19$\pm$0.20 &                                                              &      \\ 
RX J162835.4+781917 &  39 & 16 28 35.50 &78 19 17.9 &  4.3 &   306.9 & 2.12e-03 $\pm$1.6e-04 &  0.41$\pm$0.07 &  0.09$\pm$0.08 & U63298(18.3,0.0,5.0), F122\_001 AGN z=1.134                  & -0.9 \\ 
RX J162904.8+781030 &  40 & 16 29 04.83 &78 10 30.0 &  4.2 &   127.5 & 8.59e-04 $\pm$9.7e-05 &  0.58$\pm$0.14 &  0.05$\pm$0.10 &                                                              &      \\ 
RX J162910.4+780441 &  41 & 16 29 10.47 &78 04 41.0 &  1.1 &  4240.2 & 1.23e-02 $\pm$3.4e-04 & -0.99$\pm$0.00 & -0.12$\pm$0.03 & U64063(13.3,0.0,2.3), RE J1629+780 WD binary                 & -2.1 \\ 
RX J162913.3+780033 &  42 & 16 29 13.34 &78 00 33.5 &  6.2 &    48.6 & 4.69e-04 $\pm$7.6e-05 &  0.65$\pm$0.26 &  0.23$\pm$0.13 & APM(20.7,0.7,4.6)                                            & -0.8 \\ 
RX J162923.9+781304 &  43 & 16 29 23.90 &78 13 04.0 &  6.2 &    67.9 & 6.34e-04 $\pm$8.8e-05 &  0.71$\pm$0.20 &  0.41$\pm$0.12 &                                                              &      \\ 
RX J162928.5+780528 &  44 & 16 29 28.57 &78 05 28.5 &  4.7 &   123.6 & 1.58e-03 $\pm$1.3e-04 & -0.99$\pm$0.00 &  0.07$\pm$0.07 &                                                              &      \\ 
RX J162930.6+781545 &  45 & 16 29 30.69 &78 15 45.5 &  8.3 &    47.2 & 5.14e-04 $\pm$8.3e-05 &  1.00$\pm$0.00 &  0.19$\pm$0.15 &                                                              &      \\ 
RX J162936.6+775814 &  46 & 16 29 36.67 &77 58 14.9 &  5.5 &    62.7 & 5.59e-04 $\pm$8.2e-05 &  0.51$\pm$0.19 &  0.04$\pm$0.13 &                                                              &      \\ 
RX J162938.0+775711 &  47 & 16 29 38.06 &77 57 11.9 & 10.7 &    10.1 & 1.68e-04 $\pm$5.1e-05 &  1.00$\pm$0.00 &  0.27$\pm$0.25 &                                                              &      \\ 
RX J162945.0+781126 &  48 & 16 29 45.04 &78 11 26.9 &  2.0 &   861.7 & 3.28e-03 $\pm$1.8e-04 &  0.28$\pm$0.05 & -0.06$\pm$0.05 & U64758(17.7,2.0,1.0), F122\_10 Star M5.5e                    & -1.7 \\ 
RX J162956.3+781302 &  49 & 16 29 56.39 &78 13 02.7 &  3.5 &   241.0 & 1.31e-03 $\pm$1.2e-04 &  0.66$\pm$0.10 &  0.17$\pm$0.09 &                                                              &      \\ 
RX J162956.5+781705 &  50 & 16 29 56.51 &78 17 05.3 & 12.2 &    16.9 & 2.68e-04 $\pm$6.5e-05 &  1.00$\pm$0.00 & -0.02$\pm$0.25 & U65068(19.4,0.6,6.5)                                         & -1.6 \\ 
\noalign{\smallskip}\hline
\end{tabular}
\label{tab-cat}
\end{table}
\addtocounter{table}{-1}
\begin{table}
\caption[]{Continued}
\scriptsize
\begin{tabular}{rrrrrrrrrp{70mm}l}
\hline\noalign{\smallskip}
\multicolumn{1}{c}{1} & \multicolumn{1}{c}{2} & \multicolumn{1}{c}{3} &
\multicolumn{1}{c}{4} & \multicolumn{1}{c}{5} & \multicolumn{1}{c}{6} &
\multicolumn{1}{c}{7} & \multicolumn{1}{c}{8} & \multicolumn{1}{c}{9} &
\multicolumn{1}{l}{10}  & \multicolumn{1}{c}{11}\\
\hline\noalign{\smallskip}
\multicolumn{1}{c}{Source name} &\multicolumn{1}{c}{No} & \multicolumn{2}{c}{RA~~~(J2000.0)~~~Dec} &
\multicolumn{1}{c}{\perr} & \multicolumn{1}{c}{\exil} &
\multicolumn{1}{c}{Count rate} & \multicolumn{1}{c}{HR1} & \multicolumn{1}{c}{HR2} & Identifications & 
\multicolumn{1}{c}{log(${\rm f}_{\rm x} \over {\rm f}_{\rm v}$)}\\
 & & \multicolumn{1}{c}{[h m s]} & \multicolumn{1}{c}{[\degr\ \arcmin\ \arcsec]} &
\multicolumn{1}{c}{[\arcsec]} & & \multicolumn{1}{c}{[\ct]} & & & &\\
\noalign{\smallskip}\hline\noalign{\smallskip}
RX J162957.2+775746 &  51 & 16 29 57.26 &77 57 46.2 &  4.7 &   106.7 & 8.50e-04 $\pm$9.9e-05 &  0.56$\pm$0.14 & -0.04$\pm$0.10 &                                                              &      \\ 
RX J163003.8+781905 &  52 & 16 30 03.89 &78 19 05.7 &  9.7 &    53.0 & 7.65e-04 $\pm$1.1e-04 &  0.42$\pm$0.17 & -0.04$\pm$0.15 &                                                              &      \\ 
RX J163005.9+780814 &  53 & 16 30 05.95 &78 08 14.6 &  9.9 &    12.6 & 1.81e-04 $\pm$5.1e-05 &  1.00$\pm$0.00 & -0.05$\pm$0.24 &                                                              &      \\ 
RX J163016.8+781539 &  54 & 16 30 16.89 &78 15 39.0 &  8.7 &    43.9 & 4.87e-04 $\pm$8.1e-05 &  0.45$\pm$0.23 & -0.21$\pm$0.16 & U65426(17.5,0.9,5.3), U65470(18.2,-0.5,5.6)                  & -2.2 \\ 
RX J163019.9+775943 &  55 & 16 30 20.00 &77 59 43.9 &  6.0 &    44.2 & 4.38e-04 $\pm$7.4e-05 &  0.71$\pm$0.29 &  0.17$\pm$0.15 &                                                              &      \\ 
RX J163030.6+775757 &  56 & 16 30 30.69 &77 57 57.7 &  5.9 &    54.4 & 5.24e-04 $\pm$8.0e-05 &  0.15$\pm$0.15 &  0.25$\pm$0.13 & U65771(19.5,0.5,1.4), U65814(17.3,0.9,5.3)                   & -1.2 \\ 
RX J163038.4+781111 &  57 & 16 30 38.43 &78 11 11.6 &  3.6 &   285.5 & 2.00e-03 $\pm$1.5e-04 &  0.77$\pm$0.08 &  0.12$\pm$0.07 &                                                              &      \\ 
RX J163050.4+780740 &  58 & 16 30 50.46 &78 07 40.8 & 10.9 &    10.4 & 1.68e-04 $\pm$5.1e-05 &  1.00$\pm$0.00 &  0.05$\pm$0.29 & U66177(15.9,0.6,9.2)                                         & -3.2 \\ 
RX J163052.7+781636 &  59 & 16 30 52.75 &78 16 36.8 & 12.4 &    16.9 & 2.63e-04 $\pm$6.4e-05 &  1.00$\pm$0.00 & -0.20$\pm$0.26 &                                                              &      \\ 
RX J163054.5+781102 &  60 & 16 30 54.53 &78 11 02.3 &  1.6 &  1819.4 & 5.82e-03 $\pm$2.4e-04 &  0.91$\pm$0.03 &  0.40$\pm$0.04 & U66298(19.7,0.7,1.9), F122\_13 AGN z=0.358                   & -0.2 \\ 
RX J163102.6+780058 &  61 & 16 31 02.67 &78 00 58.0 & 10.3 &    18.8 & 2.70e-04 $\pm$6.4e-05 &  0.30$\pm$0.30 &  0.30$\pm$0.19 & U66448(13.3,1.4,13.9)                                        & -4.3 \\ 
RX J163105.4+782212 &  62 & 16 31 05.49 &78 22 12.0 & 18.3 &    11.4 & 3.60e-04 $\pm$9.4e-05 &  0.16$\pm$0.27 & -0.08$\pm$0.20 & U66507(18.0,1.2,16.9), U66551(18.2,-0.1,20.6)                & -2.2 \\ 
RX J163108.2+774930 &  63 & 16 31 08.30 &77 49 30.4 &  9.6 &    45.8 & 6.37e-04 $\pm$9.9e-05 &  1.00$\pm$0.00 &  0.44$\pm$0.14 & U66635(16.0,1.3,6.5)                                         & -2.8 \\ 
RX J163111.7+780028 &  64 & 16 31 11.73 &78 00 28.3 &  6.3 &    47.4 & 4.73e-04 $\pm$7.8e-05 &  0.53$\pm$0.24 &  0.45$\pm$0.15 &                                                              &      \\ 
RX J163114.9+780419 &  65 & 16 31 14.98 &78 04 19.2 &  6.9 &    40.3 & 4.65e-04 $\pm$7.8e-05 &  1.00$\pm$0.00 &  0.16$\pm$0.16 &                                                              &      \\ 
RX J163124.7+781445 &  66 & 16 31 24.70 &78 14 45.4 & 15.2 &    14.2 & 2.55e-04 $\pm$6.6e-05 &  1.00$\pm$0.00 & -0.20$\pm$0.30 & U67034(19.7,0.9,9.5)                                         & -1.6 \\ 
RX J163129.6+781642 &  67 & 16 31 29.68 &78 16 42.3 & 14.6 &    20.3 & 3.43e-04 $\pm$7.6e-05 &  1.00$\pm$0.00 &  0.20$\pm$0.25 &                                                              &      \\ 
RX J163140.9+781236 &  68 & 16 31 40.93 &78 12 36.4 &  1.9 &  1695.9 & 6.05e-03 $\pm$2.4e-04 &  0.58$\pm$0.04 &  0.08$\pm$0.04 & U67345(18.4,-0.1,1.1), F122\_14 AGN z=0.380                  & -0.3 \\ 
RX J163151.9+781756 &  69 & 16 31 51.95 &78 17 56.6 &  9.6 &    55.5 & 7.01e-04 $\pm$1.0e-04 &  0.92$\pm$0.28 &  0.22$\pm$0.15 &                                                              &      \\ 
RX J163152.7+780602 &  70 & 16 31 52.79 &78 06 03.0 &  7.0 &    31.4 & 3.43e-04 $\pm$6.7e-05 &  1.00$\pm$0.00 &  0.38$\pm$0.20 &                                                              &      \\ 
RX J163155.7+780443 &  71 & 16 31 55.72 &78 04 43.9 & 10.7 &    21.7 & 3.41e-04 $\pm$7.2e-05 &  0.56$\pm$0.32 & -0.29$\pm$0.21 &                                                              &      \\ 
RX J163157.0+775723 &  72 & 16 31 57.06 &77 57 23.3 & 11.9 &    11.9 & 2.40e-04 $\pm$6.5e-05 &  1.00$\pm$0.00 &  0.03$\pm$0.22 &                                                              &      \\ 
RX J163159.8+780958 &  73 & 16 31 59.88 &78 09 58.2 &  4.8 &   144.2 & 1.03e-03 $\pm$1.1e-04 &  1.00$\pm$0.00 &  0.19$\pm$0.11 &                                                              &      \\ 
RX J163207.3+781319 &  74 & 16 32 07.38 &78 13 19.5 &  3.4 &   398.3 & 2.07e-03 $\pm$1.5e-04 &  1.00$\pm$0.00 &  0.28$\pm$0.07 & U67910(19.7,0.7,2.6)                                         & -0.6 \\ 
RX J163213.5+781029 &  75 & 16 32 13.50 &78 10 29.7 & 10.5 &    47.6 & 6.49e-04 $\pm$9.6e-05 &  1.00$\pm$0.00 &  0.27$\pm$0.16 & U67986(19.8,0.4,5.8)                                         & -0.9 \\ 
RX J163216.6+775411 &  76 & 16 32 16.68 &77 54 11.5 &  2.0 &  1485.4 & 6.28e-03 $\pm$2.6e-04 &  0.56$\pm$0.04 &  0.00$\pm$0.04 & U68106(13.1,1.1,0.5), F122\_31 Star K0e                      & -2.9 \\ 
RX J163220.5+781640 &  77 & 16 32 20.54 &78 16 40.4 &  7.2 &    98.8 & 9.49e-04 $\pm$1.1e-04 &  1.00$\pm$0.00 &  0.40$\pm$0.11 &                                                              &      \\ 
RX J163225.4+781118 &  78 & 16 32 25.44 &78 11 18.2 &  5.2 &   423.8 & 4.26e-03 $\pm$2.1e-04 &  0.98$\pm$0.05 &  0.15$\pm$0.05 &                                                              &      \\ 
RX J163238.1+781508 &  79 & 16 32 38.13 &78 15 08.2 & 18.2 &    11.8 & 2.85e-04 $\pm$7.5e-05 &  1.00$\pm$0.00 & -0.06$\pm$0.31 & NVSS(3.0,9.9)                                                &      \\ 
RX J163238.7+781151 &  80 & 16 32 38.79 &78 11 51.1 &  1.3 &  5375.9 & 1.56e-02 $\pm$3.9e-04 &  0.89$\pm$0.02 &  0.00$\pm$0.03 & NGC 6217, F122\_16 R11.2 ELG LINER, NVSS(79.9,2.4)           & -2.8 \\ 
RX J163245.5+780620 &  81 & 16 32 45.59 &78 06 20.8 &  8.8 &    26.3 & 3.58e-04 $\pm$7.3e-05 &  0.88$\pm$0.44 &  0.30$\pm$0.20 &                                                              &      \\ 
RX J163307.1+780849 &  82 & 16 33 07.12 &78 08 49.7 & 10.9 &    16.5 & 2.74e-04 $\pm$6.7e-05 &  1.00$\pm$0.00 & -0.12$\pm$0.24 &                                                              &      \\ 
RX J163316.3+780432 &  83 & 16 33 16.34 &78 04 32.2 &  4.9 &   127.0 & 1.09e-03 $\pm$1.2e-04 &  0.89$\pm$0.15 &  0.32$\pm$0.10 &                                                              &      \\ 
RX J163329.5+780847 &  84 & 16 33 29.51 &78 08 47.9 &  9.7 &    33.3 & 4.64e-04 $\pm$8.4e-05 &  1.00$\pm$0.00 &  0.33$\pm$0.16 & U69617(19.4,0.9,8.5)                                         & -1.5 \\ 
RX J163348.0+775644 &  85 & 16 33 48.01 &77 56 44.7 & 12.4 &    21.1 & 4.14e-04 $\pm$8.7e-05 &  0.73$\pm$0.41 &  0.55$\pm$0.22 &                                                              &      \\ 
RX J163354.5+775329 &  86 & 16 33 54.56 &77 53 29.7 & 10.1 &    62.2 & 8.45e-04 $\pm$1.1e-04 &  0.57$\pm$0.19 & -0.08$\pm$0.16 &                                                              &      \\ 
RX J163423.1+775651 &  87 & 16 34 23.18 &77 56 51.4 &  5.6 &   207.1 & 1.87e-03 $\pm$1.6e-04 &  0.56$\pm$0.10 &  0.06$\pm$0.09 & U70795(18.2,-0.3,8.1)                                        & -0.8 \\ 
RX J163425.5+781002 &  88 & 16 34 25.57 &78 10 02.9 &  3.5 &   636.9 & 3.93e-03 $\pm$2.1e-04 &  1.00$\pm$0.00 &  0.42$\pm$0.05 & F122\_21 R19.2 AGN z=0.376                                   & -0.2 \\ 
RX J163432.5+775334 &  89 & 16 34 32.58 &77 53 34.1 & 10.6 &    80.9 & 1.22e-03 $\pm$1.4e-04 &  0.50$\pm$0.13 &  0.18$\pm$0.14 &                                                              &      \\ 
RX J163439.6+780127 &  90 & 16 34 39.65 &78 01 27.7 & 12.1 &    25.5 & 4.81e-04 $\pm$9.3e-05 &  1.00$\pm$0.00 &  0.28$\pm$0.16 &                                                              &      \\ 
RX J163503.3+780420 &  91 & 16 35 03.39 &78 04 20.5 & 11.2 &    47.7 & 7.96e-04 $\pm$1.2e-04 &  0.63$\pm$0.21 &  0.08$\pm$0.12 &                                                              &      \\ 
\noalign{\smallskip}\hline\noalign{\smallskip}
\end{tabular}

References:\\
Un: source 1650-018n from USNO-A2 Catalogue \citep{1998USNO2.C......0M}; 
Fn: \citet{2000MNRAS.311..456M}; 
WN Bn: source in Westerbork Northern Sky Survey \citep[WENSS,][]{1997A&AS..124..259R}; 
NVSS: \citet{1998AJ....115.1693C}; 
APM: \citet{1994Spec....2...14I}

\end{table}

\end{document}